\documentclass[a4paper,scale=0.8,onecolumn,nobibnotes,nofootinbib]{revtex4}
\usepackage{enumitem}
\usepackage{amsmath,amssymb}
\usepackage{graphicx} 
\usepackage{epstopdf}
\usepackage{hyperref}
\usepackage{geometry}
\hypersetup{
    colorlinks=true,
    citecolor=blue,
    linkcolor=red,
    filecolor=magenta,
    urlcolor=cyan,}

\newcommand{\be}{\begin{equation}}
\newcommand{\ee}{\end{equation}}

\begin{document}

\title{Thermodynamics of the Reissner-Nordstr\"om black hole with quintessence matter on the EGUP framework}
\author{Hao Chen\footnote{gs.ch19@gzu.edu.cn}$^{1}$, Bekir Can L\"{u}tf\"{u}o\u{g}lu\footnote{bclutfuoglu@akdeniz.edu.tr.}$^{2,3}$, Hassan Hassanabadi\footnote{h.hasanabadi@shahroodut.ac.ir}$^{3,4}$, and Zheng-Wen Long\footnote{zwlong@gzu.edu.cn (Corresponding author)}$^{1}$}
\affiliation{$^1$ College of Physics, Guizhou University, Guiyang, 550025, China.\\
$^{2}$ Department of Physics, Akdeniz University, Campus, 07058 Antalya, Turkey.\\
  $^{3}$ Department of Physics, University of Hradec Kr$\acute{a}$lov$\acute{e}$, Rokitansk$\acute{e}$ho 62, 500 03 Hradec Kr$\acute{a}$lov$\acute{e}$, Czech Republic. \\
$^{4}$ Faculty of Physics, Shahrood University of Technology, Shahrood, Iran.
}

\date{\today}

\begin{abstract}
In this paper, we investigate the quantum correction on thermodynamics
of the Reissner-Nordstr\"om black hole in the presence of the quintessence matter associated with dark energy. To this end, the modified Hawking temperature, heat capacity, and entropy functions of the black hole are derived. { Investigation reveals that the modified uncertainty principles and normalization factor $\alpha$ restrict the lower bound value of horizon radius to affect the Hawking temperature and the parameter $\eta_{0}$ have a non-negligible effect on the heat capacity threshold.} {Furthermore}, the correction of entropy does not depend on the quintessence matter. It is also examined the equation of state associated with the pressure and the volume in this framework. In addition, we employ graphical methods to compare and discuss the findings within the generalized uncertainty principle (GUP) and extended uncertainty principle (EUP).
\end{abstract}
\maketitle
~~~~~~~~~\textsl{Keywords:}  Dark energy, Reissner-Nordstr\"om black hole, Thermodynamics, Extended generalized uncertainty principle.

\section{Introduction}

Over the last decade, black holes have been recognized as one of the most fascinating topics in physics. This fact is based on Bekenstein's article \cite{in1}, which suggested that black holes can be regarded as physical systems rather than mathematical constructions, and thus their physics can be interpreted within the framework of information theory. This approach leads to the integration of the classical theory of gravity, and mechanics with the quantum theory in the framework of black hole thermodynamics to investigate quantum gravity \cite{king1}. Recent astronomical observations indicate the existence of the accelerated expansion of the universe \cite{king2}, which can be explained by the string theory, as well as the concept of dark energy. For example, within the string cloud model that is constituted with the classical relativistic strings \cite{ST1}, { one can investigate black hole solution and thermodynamics in the Einstein-Gauss-Bonnet and Lovelock gravities \cite{ST2,ST4, ST5}.} Very recently, Cai et al. investigated the quasi-normal modes of a Schwarzschild black hole embedded in a string cloud in Rastall gravity \cite{ST3}. The quintessence matter (QM) approach  is a scalar field-based model to describe dark energy \cite{qui2,qui3,Copeland2006}. Kiselev, in Ref. \cite{qui1} presented the symmetrically exact solutions of the Einstein's equation for black holes surrounded by QM, in which, the value of the barotropic index, $\omega_{q}$, plays a role in the formation of the black hole horizon in the accelerated expansion scenario of the universe \cite{,quii1}. {The thermodynamical properties of black holes are shown to be affected by QM.} For example, authors in Ref. \cite{chen1} discussed such an influence by examining the Hawking radiation of a static d-dimensional spherically symmetric black hole in the presence of QM. Recently, some authors investigated the thermodynamics of the black holes with QM background \cite{da1,da2, B6}.

The existence of a minimal physical length concept has been confirmed in many fields of physics, such as quantum gravity \cite{da3}, the string theory \cite{da4}, and the non-commutative geometry \cite{da5}. In the framework of quantum mechanics, a minimal observable length concept enforces a modification in the usual Heisenberg algebra and  uncertainty principle \cite{k1, Kempf95}. The latter is referred  to as the generalized uncertainty principle (GUP).  In literature, we observe various modified  Heisenberg  algebras which present different physical consequences \cite{EUP1, B1, ali1, Pedram2012,chung1,hh1,zhao1}. Modified Heisenberg algebra is frequently used to examine black hole thermodynamics as well as to investigate quantum mechanical processes. For example, in \cite{Zhu1,Luciano2021,B2,B3,B4,B5,B7,uf1} authors discussed the Unruh effect,  Hawking temperature, mass-temperature, entropy, and specific heat functions of black holes. They showed that in some scenarios, black holes with masses less than a certain minimum mass value cannot exist. Moreover, {they discussed the conditions where the black holes stop
evaporating and leave remnant masses behind.} It is worth noting that the thermodynamics of the Schwarzschild black hole surrounded by {QM is investigated} within the GUP formalism  very recently in \cite{B6}.

Inspired by the minimum physical length \cite{Kempf95}, Kowalski-Glikman and Smolin considered a new algebra in de Sitter space-time and proposed a correction term that is proportional to the radius of space-time, to characterize a minimum physical momentum value \cite{EUP1}. Later, this modification is called the extended uncertainty principle (EUP). Recently, based on the investigation of the weak deflection angle of the black holes \cite{uf2,uf3,uf4,uf6}, the authors considered the EUP to study the effects of quantum fluctuations spewed by a black hole \cite{uf7}. Besides, Chung and Hassanabadi employed EUP to investigate the thermodynamic properties of the black hole \cite{hass1}. More interestingly, Bolen and Cavagli\'a
combined the GUP and EUP formalism, they considered the EGUP to investigate the Schwarzschild (anti-)de Sitter black hole thermodynamics \cite{tt2}. In 2009, Zhu et al. explored the influence of the EGUP formalism on the thermodynamics of the Friedmann-Robertson-Walker universe \cite{f4}. Very recently,  Hassanabadi et al. have used the EGUP to discuss the Unruh effects of a black hole \cite{tt3}.

Until now, the thermodynamics of various black holes have been extensively investigated through many modified uncertainty principles. In this manuscript, we would like to analyze the thermodynamics of the Reissner-Nordstr\"om black hole which is surrounded by QM via the EGUP formalism in the large length and  high energy scales. To this end, we construct the manuscript as follows: In section \ref{sec2}, we briefly introduce the EGUP formalism. Then, in section \ref{sec3}, we derive and analyze the thermodynamic properties of the Reissner-Nordstr\"om black hole with the considered formalism in the presence of  QM. After analyzing the results in the GUP and EUP limits, we conclude the manuscript in section \ref{sec:conclusion}.

\section{A brief review of the modified formalism} \label{sec2}
According to the quantum theory, black hole radiation has to be taken as a quantum effect. Therefore, particles have to satisfy the Heisenberg uncertainty principle:
{\begin{equation}
\Delta x_{i} \Delta p_{j} \geq \frac{\hbar }{2}\delta_{ij}, \quad \quad i, j=1, 2, 3.
\end{equation}}
Where $x_{i}$ and $p_{j}$ are the spatial coordinate and momentum operators, respectively, while $\delta_{ij}$ is the Kronecker delta, and $\hbar$ denotes the reduced Planck constant. However, a minimal physical length concept does not arise from the usual Heisenberg algebra, instead, it can appear with a modified algebra \cite{k1} \begin{equation}\label{P1}
\Delta x_{i} \Delta p_{j} \geq \frac{\hbar }{2} \delta_{i j}\left[1+\frac{\beta_0 l_{p}^{2}}{\hbar^{2}}\left(\Delta p\right)^{2} \right].
\end{equation}
{Here, $\beta_0$ is the dimensionless GUP parameter and is of order unity, while $l_p$ is the Planck length.} In this case, an observable minimal length is found to be $\Delta x_{i}\geq \sqrt{\beta_0} l_{p}$. Furthermore,  Eq. (\ref{P1}) leads to an uncertainty range of momentum:
\begin{equation}
\frac{\hbar \Delta x_{j}}{\beta_0 l_{p}^{2}}\left[1-\sqrt{1-\frac{ \beta_0 l_{p}^{2}}{\left(\Delta x\right)^{2}}}\right] \leq \Delta p_{j} \leq \frac{\hbar \Delta x_{j}}{\beta_0 l_{p}^{2}}\left[1+\sqrt{1-\frac{ \beta_0 l_{p}^{2}}{\left(\Delta x\right)^{2}}}\right].
\end{equation}

It is worth noting that, the correction term contributes to the uncertainty meaningfully if and only if the { position uncertainty is of the order of the Planck length.} Therefore, such a scenario can be expected only in the very early stages of the universe. Recently, Hassanabadi and Chung proposed a new GUP formalism in the form of
\cite{hh1}:
\begin{equation}
\Delta x_{i} \Delta p_{j} \geq \frac{\hbar \delta_{i j}}{2}\left[-\beta(\Delta p_{j})+\frac{1}{1-\beta(\Delta p_{j})}\right],
\end{equation}
where $\beta$ is associated with the dimensionless GUP parameter. This new GUP formalism leads to a maximal momentum bound value,  $1/ \beta$, as well as a minimal length value, $3 \beta \hbar /2$. In Ref. \cite{zhao1}, the authors proposed a new high-order GUP with a minimal uncertainty of position, $\frac{3 \sqrt{3}}{4} \hbar \sqrt{\beta}$, in the following form
\begin{equation}
\Delta x_i \Delta p_j \geq \frac{\hbar\delta_{ij}}{2} \frac{1}{1-\beta\left[(\Delta p)^{2}-\langle p\rangle^{2}\right]}.
\end{equation}
Alike, a minimal measurable momentum concept is proposed in the quantum-mechanical systems \cite{Kempf95}. In other words, the ordinary Heisenberg uncertainty principle is modified as \cite{EUP1}
\begin{equation}
\Delta x_{i} \Delta p_{j} \geq \frac{\hbar\delta_{ij}}{2}\bigg[1+\eta^{2} (\Delta x)^{2}\bigg],
\end{equation}
here the deformation parameter satisfies $\eta^{2}=1 / l_{H}^{2}$, while $l_{H}$ denotes the (anti-)de Sitter space-time radius. In this formalism, we find a minimal momentum value as $(\triangle p)_{\min }=\hbar/ l_{H}$. It is worth noting that, unlike the GUP, the EUP formalism is believed to play an important role in the latter stages of the universe.

In this manuscript, we consider the following EGUP formalism as a linear combination of the GUP and EUP formalism,
\begin{equation}\label{kq}
\Delta x_{i} \Delta p_{j} \geq \frac{\hbar \delta_{i j}}{2}\left[1+\beta_0 l_{p}^{2} \frac{\left(\Delta p\right)^{2}}{\hbar^{2}}+\eta_0 \frac{\left(\Delta x\right)^{2}}{l^{2}}\right],
\end{equation}
where $\beta_0$ and $\eta_0$  are positive deformation parameters that depend on the expectation values of $x$ and $p$. Here, $l$ is a large-scale quantity associated with space-time radius. In this case, Eq. (\ref{kq}) leads to the following uncertainty ranges of momentum
\begin{equation} \label{pp3}
\frac{\hbar \Delta x_{i}}{\beta_0 l_{p}^{2}}\left[1-\sqrt{1-\frac{\beta_0 l_{p}^{2}}{\left(\Delta x\right)^{2}}\left(1+\frac{\eta_0\left(\Delta x\right)^{2}}{l^{2}}\right)}\right] \leq \Delta p_{i} \leq \frac{\hbar \Delta x_{i}}{\beta_0 l_{p}^{2}}\Bigg[1+\sqrt{1-\frac{\beta_0 l_{p}^{2}}{\left(\Delta x\right)^{2}}\left(1+\frac{\eta_0\left(\Delta x\right)^{2}}{l^{2}}\right)} \Bigg],
\end{equation}
and position
\small
\begin{equation}
\frac{l^{2} \Delta p_{i}}{\hbar \eta_0}\Bigg[1-\sqrt{1-\frac{\eta_0 \hbar^{2}}{l^{2}\left(\Delta p\right)^{2}}\left(1+\frac{\beta_0 l_{p}^{2}\left(\Delta p\right)^{2}}{\hbar^{2}}\right)}\Bigg] \leq \Delta x_{i} \leq \frac{l^{2} \Delta p_{i}}{\hbar \eta_0}\Bigg[1+\sqrt{1-\frac{\eta_0 \hbar^{2}}{l^{2}\left(\Delta p\right)^{2}}\left(1+\frac{\beta_0 l_{p}^{2}\left(\Delta p\right)^{2}}{\hbar^{2}}\right)}\Bigg].
\end{equation}\normalsize
It is worth emphasizing that the EGUP formalism leads to minimum position and momentum physical values simultaneously. In the next section, we employ this formalism and investigate its influence on the thermodynamics of the Reissner-Nordstr\"om black hole surrounded by QM. Throughout the rest of the manuscript, we use natural units.

\section{Thermodynamics of the Reissner-Nordstr\"om black hole embedded in QM within the EGUP formalism} \label{sec3}
One of the basic motivations of our manuscript originates from {Kiselev's} work \cite{qui1}. Based on the solution of Einstein's field equations associated with a static spherically symmetric black hole with QM, the general form of the considered metric reads
\begin{equation}
d s^{2}=-g(r) d t^{2}+\frac{1}{g(r)} d r^{2}+r^{2} d \theta^{2}+r^{2} \sin ^{2} \theta d \phi^{2},
\end{equation}
where $g(r)$ is given by
\begin{equation}\label{pqt1}
g(r)=1-\frac{2 M}{r}+\frac{Q^{2}}{r^{2}}-\frac{\alpha}{r^{3 \omega_{q}+1}},
\end{equation}
for the  Reissner-Nordstr\"om black hole with the mass, $M$.
Here, $Q$ is the charge of the black hole, and $\alpha$ is a positive normalization factor associated with QM. Regarding the accelerated expansion of the universe the barotropic index can take values in the range, $-1<\omega_{q}<-1/3$. One can find some typical scenarios  based on the values of the barotropic index in \cite{quii1}.
For example, the free QM
generates the black hole horizon and the (anti-)de Sitter radius for $-1<\omega_{q}<0$ and $-1<\omega_{q}<-2/3$, respectively. Moreover, $\omega_{q}=-1$, and $\omega_{q}=-2/3$ are fixed for the cosmological constant and quintessence {regime of dark energy, respectively.} We obtain the ordinary Reissner-Nordstr\"om black hole if take the normalization factor parameter to satisfy $\alpha=0$. Furthermore, if we also ignore the electric charge, $Q=0$, then we arrive at the standard Schwarzschild black hole.

We start by deriving the event horizon out of Eq. (\ref{pqt1}) via
\begin{equation}\label{t1}
\bigg(1-\frac{2 M}{r}+\frac{Q^{2}}{r^{2}}-\frac{\alpha}{r^{3 \omega_{q}+1}}\bigg)\bigg|_{r=r_{h}}=0.
\end{equation}
Then, we define the Hawking temperature, $T$, in the semi-classical case and assume the entropy, $S$, as a function of the black hole area, $A$, \cite{k7}
\begin{equation} \label{pp20}
T=\frac{\kappa}{8 \pi} \frac{d A}{d S},
\end{equation}
where $\kappa$ is the surface gravity that can be calculated via
\begin{equation}\label{pp1}
\kappa=-\lim _{r \rightarrow r_{h}} \sqrt{-\frac{g^{11}}{g^{00}}} \frac{d\left(\ln\left(g^{00}\right)\right)}{d r}=\frac{1}{r_{h}}\left(1-\frac{Q^{2}}{r_{h}^{2}}+\frac{3 \omega_{q}\alpha}{r_{h}^{3 \omega_{q}+1}}\right).
\end{equation}
Within a heuristic approach \cite{k7}, we assume that the black hole absorbs a  particle that is located in the vicinity of its horizon. The process results in a minimal increase in the black hole area that is proportional to the product of the position and momentum uncertainties. {Similarly, Medved and Vagenas discussed the quantum corrections of entropy, area and the calibration factor by employing the generalized uncertainty principle \cite{Medved1}}. Also, the process leads to a minimal change in the entropy which can not be smaller than $\ln 2$. Therefore, we write
\begin{equation} \label{pp2}
T \simeq \frac{\kappa(\Delta A)_{\min }}{8 \pi(\Delta S)_{\min }} \simeq \frac{\kappa \epsilon}{8 \pi \ln 2} \Delta x \Delta p.
\end{equation}
Here, $\epsilon$ is the calibration factor that ensures the consistency of the obtained result with the semi-classical result. Based on this approach, we employ $\Delta x\simeq2r_{h}$, and  substitute Eqs. (\ref{pp3}) and (\ref{pp1}) in Eq. (\ref{pp2}). We get the modified Hawking temperature as
\begin{equation}\label{t2}
T=\frac{\epsilon r_{h}}{2 \pi \beta_0 l_{p}^{2} \ln 2}\left(1-\frac{Q^{2}}{r_{h}^{2}}+\frac{3 \alpha \omega_{q}}{r_{h}^{3 \omega_{q}+1}}\right)\left[1-\sqrt{1-\frac{\beta_0 l_{p}^{2}}{4 r_{h}^{2}}\left(1+\frac{4 \eta_0 r_{h}^{2}}{l^{2}}\right)}\right].
\end{equation}
In the absence of the QM, charge and EGUP parameters ($\alpha=Q=\beta_0=\eta_0=0$),  Eq. \eqref{t2} reduces to  $T=\epsilon /({16 \pi r_{h} \ln 2})$. To reproduce the well-known Hawking temperature, $T=1 / ({4\pi r_{h}})$ \cite{in1,k8}, we tune the calibration factor to  $\epsilon = 4 \ln 2 $. After all, we express the EGUP modified Hawking temperature associated with the Reissner-Nordstr\"om black hole surrounded with the QM in the form of
\begin{equation}\label{Te1}
T_{EGUP}=\frac{2r_{h}}{ \pi \beta_0 l_{p}^{2} }\left(1-\frac{Q^{2}}{r_{h}^{2}}+\frac{3 \alpha \omega_{q}}{r_{h}^{3 \omega_{q}+1}}\right)\left[1-\sqrt{1-\frac{\beta_0 l_{p}^{2}}{4 r_{h}^{2}}\left(1+\frac{4 \eta_0 r_{h}^{2}}{l^{2}}\right)}\right].
\end{equation}
For $\eta_0=0$, we get the GUP modified Hawking temperature
\begin{equation}\label{TGUP}
T_{GUP}=\frac{2r_{h}}{ \pi \beta_0 l_{p}^{2} }\left(1-\frac{Q^{2}}{r_{h}^{2}}+\frac{3 \alpha \omega_{q}}{r_{h}^{3 \omega_{q}+1}}\right)\left[1-\sqrt{1-\frac{\beta_0 l_{p}^{2}}{4 r_{h}^{2}}}\right].
\end{equation}
Similarly, the EUP modified Hawking temperature arises for $\beta_0=0$
\begin{equation}\label{oo1}
T_{EUP}=\frac{1}{4 \pi r_{h}}\left(1-\frac{Q^{2}}{r_{h}^{2}}+\frac{3\alpha \omega_{q}}{r_{h}^{3 \omega_{q}+1}}\right)\left(1+\frac{4 \eta_0 r_{h}^{2}}{l^{2}}\right),
\end{equation}
while for $\beta_0=0$ and $\eta_0=0$, Eq. \eqref{Te1} reduces to HUP-modified Hawking temperature
\begin{equation}
T_{HUP}=\frac{1}{4\pi r_{h}}\left(1-\frac{Q^{2}}{r_{h}^{2}}+\frac{3 \alpha\omega_{q}}{r_{h}^{3 \omega_{q}+1}}\right).
\end{equation}
It is worth noting that the Hawking temperature must be  positive and real-valued. Therefore, there can be two constraints maximally. Roughly speaking, we can say that the term originating from QM, which depends on the charge, barotropic index, and normalization factor parameters, guarantees the positive values, while the other terms with the EGUP parameters in Eq. \eqref{Te1} ensures to take real values. There is no doubt that without particular values of the charge, barotropic index, and normalization factor parameters one cannot exactly present an analysis on positively-valued Hawking temperature. For example, if we consider the simplest case and take  barotropic index as  $\omega_{q}=-\frac{1}{3}$, then the following constraint on the radius arises
\begin{equation}
1-\alpha-\frac{Q^{2}}{r^{2}_{h}}\geq0\Rightarrow r_{h} \geq |Q| \sqrt{\frac{1}{(1-\alpha)}}, \label{QMcons13}
\end{equation}
with $0 \leq \alpha \leq 1$. As another case, if we consider $\omega_{q}=-\frac{2}{3}$, then we have a third-order equation \begin{equation}
1-\frac{Q^{2}}{r^{2}_{h}}-2\alpha r_h \geq0,  \label{QMcons23}
\end{equation}
with the following roots:
\small
\begin{eqnarray}
r_{{h}_{1}}&=&\frac{1}{6 \alpha
   }-\frac{\sqrt[3]{108 \alpha
   ^2
   {Q}^2+\sqrt{\left(108
   \alpha ^2
   Q^2-2\right)^2-4}-2}
   }{6 \sqrt[3]{2} \alpha
   }-\frac{1}{3\ 2^{2/3} \alpha
    \sqrt[3]{108 \alpha ^2
   Q^2+\sqrt{\left(108
   \alpha ^2
   Q^2-2\right)^2-4}-2}
   }, \\
r_{{h}_{2}}&=&\frac{1}{6 \alpha
   }+\frac{\left(1-i
   \sqrt{3}\right) \sqrt[3]{108
   \alpha ^2
   {Q}^2+\sqrt{\left(108
   \alpha ^2
   {Q}^2-2\right)^2-4}-2}
   }{12 \sqrt[3]{2} \alpha
   }+\frac{1+i \sqrt{3}}{6\
   2^{2/3} \alpha  \sqrt[3]{108
   \alpha ^2
   {Q}^2+\sqrt{\left(108
   \alpha ^2
   {Q}^2-2\right)^2-4}-2}
   }, \nonumber \\
r_{{h}_{3}}&=&\frac{1}{6 \alpha
   }+\frac{\left(1+i
   \sqrt{3}\right) \sqrt[3]{108
   \alpha ^2
   {Q}^2+\sqrt{\left(108
   \alpha ^2
   {Q}^2-2\right)^2-4}-2}
   }{12 \sqrt[3]{2} \alpha
   }+\frac{1-i \sqrt{3}}{6\
   2^{2/3} \alpha  \sqrt[3]{108
   \alpha ^2
   {Q}^2+\sqrt{\left(108
   \alpha ^2
   {Q}^2-2\right)^2-4}-2}
   }. \nonumber
\end{eqnarray}
\normalsize
On the other hand, we can execute an algebraic analysis on the second constraint, which ensures real values to the Hawking temperature,  by examining the square root term in square brackets of Eq. \eqref{Te1}. In the EGUP scenario, we  have
\begin{equation}
1-\frac{\beta_0 l_{p}^{2}}{4 r_{h}^{2}}\left(1+\frac{4 \eta_0 r_{h}^{2}}{l^{2}}\right)\geq0,
\end{equation}
that leads to a lower bound on the horizon
\begin{equation}\label{horegup1}
 r_{h_{EGUP}} \geq \frac{l_{p}l}{2} \sqrt{\frac{\beta_0}{(l^{2}-\beta_0 \eta_0 l^{2}_{p})}}.
\end{equation}
In the GUP scenario, the constraint on the horizon radius reduces to
\begin{equation}
r_{h_{GUP}}\geq \frac{l_{p}\sqrt{\beta_0}}{2},
\end{equation}
while in the EUP and HUP scenarios, the constraint vanishes and we have $r_h\geq 0$.

To illustrate this analysis concretely, we depict the EGUP  modified Hawking temperature versus horizon for $\omega_q=-1/3$ and $\omega_q=-2/3$ in Fig. 1 and Fig. 2, respectively. In this manuscript, in all plots we will use the following values: $l_p=l=1.0$, $Q=0.3$, $\alpha=0.10$. Since the first constraint depends on the values of the charge and normalization factors, we get the following conditions:
\begin{enumerate}[label=\roman*.]
\item For $\omega_q=-1/3$ case, Eq. \eqref{QMcons13} gives $r_h \geq 0.316$.

\item For $\omega_q=-2/3$ case,  Eq. \eqref{QMcons23} restricts the horizon to be in the interval, $0.310\leq r_h \leq 4.982$.

\end{enumerate}
\begin{figure}
\begin{tabular}{cc}
\begin{minipage}[t]{0.45\linewidth}
\centerline{\includegraphics[width=7.0cm]{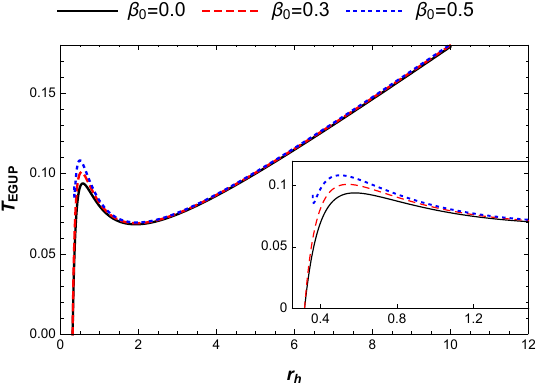}} \label{fig1a}
\centerline{(a)}
\end{minipage}
\begin{minipage}[t]{0.45\linewidth}
\centerline{\includegraphics[width=7.0cm]{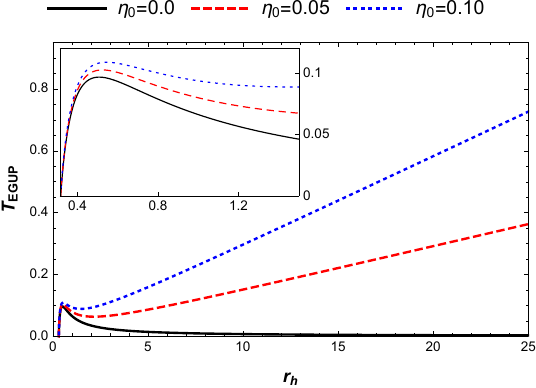}}\label{fig1b}
\centerline{(b)}
\end{minipage}
\end{tabular}
\parbox[c]{15.0cm}{\footnotesize{\bf Fig.~1.} (a)Hawking temperature $T_{-\frac{1}{3}}$ as a function of horizon radius $r_{h}$ for different GUP parameters, $\beta_0$. $(l=l_{p}=1, Q=0.3, \omega_{q}=-1/3, \alpha=0.1, \eta_0=0.06)$.\\
(b)Hawking temperature $T_{-\frac{1}{3}}$ as a function of horizon radius $r_{h}$ for different EUP parameters, $\eta_0$. $(l=l_{p}=1, Q=0.3, \omega_{q}=-1/3, \alpha=0.1, \beta_0=0.39 )$.}
\end{figure}
To observe the effect of the GUP parameter, we consider a constant EUP parameter, $\eta_0=0.06$, and three different GUP
parameters in Fig. 1(a). According to the second constraint given in Eq. \eqref{horegup1}, we find the followings:
\begin{enumerate}[label=\roman*.]

\item For $\beta_0=0.0$,  $r_h \geq 0$.

\item For $\beta_0=0.3$,  $r_h \geq 0.276$.

\item For $\beta_0=0.5$,  $r_h \geq 0.359$.
\end{enumerate}
Since there are two conditions, the solutions must be in their intersection range. Therefore, we conclude that,
\begin{enumerate}[label=\roman*.]

\item For $\beta_0=0.0$, and $\eta_0=0.06$,  $r_h \geq 0.316$ and $T_{EGUP}(0.316)=0$.

\item For $\beta_0=0.3$,  and $\eta_0=0.06$,  $r_h \geq 0.316$ and $T_{EGUP}(0.316)=0$.

\item For $\beta_0=0.5$, and $\eta_0=0.06$,  $r_h \geq 0.359$ and $T_{EGUP}(0.359)=0.092$.
\end{enumerate}
We observe that the inset plot of Fig. 1(a) confirms our prediction. Moreover, we confirm that the GUP scenario is effective in the early stage.

Then, to demonstrate the effect of the EUP parameter, we consider a constant GUP parameter, $\beta_0=0.39$, and three different EUP parameters in Fig. 1(b). According to the second constraint, we obtain the followings:
\begin{enumerate}[label=\roman*.]

\item For $\eta_0=0.00$,  $r_h \geq 0.312$.

\item For $\eta_0=0.05$,  $r_h \geq 0.315$.

\item For $\eta_0=0.10$,  $r_h \geq 0.316$.
\end{enumerate}
As we discussed above, the solutions must be in the intersection range. Therefore, we conclude that,
\begin{enumerate}[label=\roman*.]

\item For $\beta_0=0.39$, and $\eta_0=0.00$,  $r_h \geq 0.316$ and $T_{EGUP}(0.316)=0$.

\item For $\beta_0=0.39$,  and $\eta_0=0.05$,  $r_h \geq 0.316$ and $T_{EGUP}(0.316)=0$.

\item For $\beta_0=0.39$, and $\eta_0=0.10$,  $r_h \geq 0.359$ and $T_{EGUP}(0.318)=0.007$.
\end{enumerate}
The inset plot of Fig. 1(b) verifies our findings. Also, we notice that the EUP scenario becomes effective in the late stages as being discussed in the literature.

We repeat a similar analysis for $\omega_q=-2/3$  QM scenario. For $\eta_0=0.06$, we consider $\beta=0.0$, $\beta=0.3$, and $\beta=0.6$ to investigate the role of the GUP parameter. In this case, according to the second constraint, the horizon radius must be greater than $0$, $0.276$, and $0.359$, respectively. If we take the first constraint into account, we find
\begin{table}[tbh!]
    \centering
    \begin{tabular}{|c|c|c|c|c|}
    \hline
        $\beta_0$ & $\eta_0$ & horizon range & $T(r_{h_{1}})$ & $T(r_{h_{2}})$ \\
        \hline
         0.0 & 0.06 & $0.310\leq r_h \leq 4.982$ & 0 & 0 \\ \hline 0.3 & 0.06 & $0.310\leq r_h \leq 4.982$ & 0 & 0 \\\hline
         0.5 & 0.06 & $0.359\leq r_h \leq 4.982$ & 0.105 & 0 \\ \hline
    \end{tabular}
    \caption{The effect of GUP parameter in the quintessence regime of dark energy. }
    \label{tab:tab0}
\end{table}

In Fig. 2(a), we depict the EGUP modified Hawking temperature versus horizon. The inset plot confirms the results given in Table \ref{tab:tab1}. Then, we explore the role of the EUP parameter.
We take $\beta_0=0.38$, with $\eta_0=0$, $\eta_0=0.1$, and $\eta_0=0.2$. In this case, the second constraint gives the following bound values $r \geq 0.308$, $r \geq 0.311$, $r \geq 0.321$, respectively. Therefore, we can summarize the characteristic of the horizon range and Hawking temperature within Table 2.
\begin{table}[tbh!]
    \centering
    \begin{tabular}{|c|c|c|c|c|}
    \hline
        $\beta_0$ & $\eta_0$ & horizon range & $T(r_{h_{1}})$ & $T(r_{h_{2}})$ \\
        \hline
         0.38 & 0.0 & $0.310\leq r_h \leq 4.982$ & 0 & 0 \\ \hline
         0.38 & 0.1 & $0.311\leq r_h \leq 4.982$ & 0.014 & 0 \\\hline
         0.38 & 0.2 & $0.314\leq r_h \leq 4.982$ & 0.033 & 0 \\ \hline
    \end{tabular}
    \caption{The effect of EUP parameter in the quintessence regime of dark energy. }
    \label{tab:tab1}
\end{table}
In Fig. 2(b), we show how the EUP parameter affects the modified Hawking temperature within the horizon range.

\begin{figure}
\begin{tabular}{cc}
\begin{minipage}[t]{0.45\linewidth}
\centerline{\includegraphics[width=7.0cm]{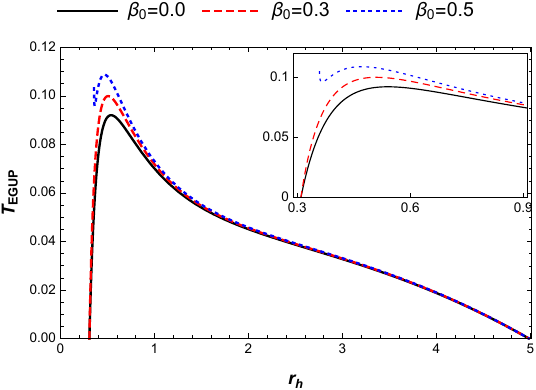}}
\centerline{(a)}
\end{minipage}
\begin{minipage}[t]{0.45\linewidth}
\centerline{\includegraphics[width=7.0cm]{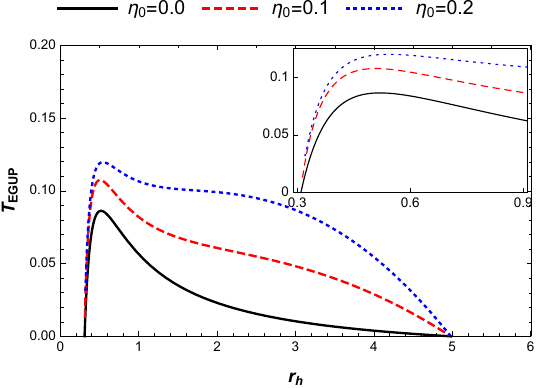}}
\centerline{(b)}
\end{minipage}
\end{tabular}
\parbox[c]{15.0cm}{\footnotesize{\bf Fig.~2.} (a)Hawking temperature $T_{-\frac{2}{3}}$ as a function of horizon radius $r_{h}$ for different GUP parameters $\beta_0$ $(l=l_{p}=1, Q=0.3, \omega_{q}=-2/3, \alpha=0.1, \eta_0=0.06)$.\\
(b)Hawking temperature $T_{-\frac{2}{3}}$ as a function of horizon radius $r_{h}$ for different EUP parameters $\eta_0$ $(l=l_{p}=1, Q=0.3, \omega_{q}=-2/3, \alpha=0.1, \beta_0=0.38 )$.}
\end{figure}

Next, we derive the EGUP modified heat capacity function from the  black hole thermodynamics \cite{hass1}
\begin{equation}\label{oo2}
C_{EGUP}=\frac{dM}{dT}=\frac{\pi \beta_0 l_{p}^{2}}{4\left(\Xi_{1}+\Xi_{2}\right)}\left(1-\frac{Q^{2}}{r_{h}^{2}}+\frac{3 \alpha \omega_{q}}{r_{h}^{3 \omega_{q}+1}}\right) \sqrt{1-\frac{\beta_0 l_{p}^{2}}{4 r_{h}^{2}}\left(1+\frac{4 \eta_0 r_{h}^{2}}{l^{2}}\right)},
\end{equation}
where
\begin{eqnarray}
\Xi_1&=& \left(1+\frac{Q^{2}}{r_{h}^{2}}-\frac{9\alpha \omega_q^2}{r_h^{3\omega_q+1}}\right)\left[\sqrt{1-\frac{\beta_0 l_{p}^{2}}{4 r_{h}^{2}}\left(1+\frac{4 \eta_0 r_{h}^{2}}{l^{2}}\right)}-1+\frac{\beta_0 l_{p}^{2}}{4 r_{h}^{2}}\left(1+\frac{4 \eta_0 r_{h}^{2}}{l^{2}}\right)\right],\\
\Xi_2&=&-\frac{\beta_0 l_{p}^{2}}{4 r_{h}^{2}}\left(1-\frac{Q^{2}}{r_{h}^{2}}+\frac{3 \alpha \omega_{q}}{r_{h}^{3 \omega_{q}+1}}\right).
\end{eqnarray}
When the heat capacity is zero, the black hole cannot exchange radiation with its surrounding space, and a black hole remnant is formed. In the case we examined, a black hole remnant occurs for $r_{h} = \frac{l_{p}l}{2} \sqrt{\frac{\beta_0}{(l^{2}-\beta_0 \eta_0 l^{2}_{p})}}$. Accordingly, we find the remnant temperature
\begin{equation}
 T_{{rem}-{EGUP}}= \frac{l}{\pi l_{p} \sqrt{\beta_0(l^{2}-\beta_0 \eta_0 l^{2}_{p})}}\left[1-\frac{4 Q^{2}(l^{2}-\beta_0 \eta_0 l^{2}_{p})}{\beta_0 l^{2}l^{2}_{p}}+3 \omega_{\mathrm{q}} \alpha \left(\frac{4(l^{2}-\beta_0 \eta_0 l^{2}_{p})}{\beta_0 l^{2}l^{2}_{p}}\right)^{\frac{3 \omega_{\mathrm{q}}+1}{2}}\right],
\end{equation}
and remnant mass
\begin{equation}
M_{{rem}-{EGUP}}=\frac{1}{2}\left[\frac{l l_{p}}{2} \sqrt{\frac{\beta_0}{\left(l^{2}-\eta_0 \beta_0 l_{p}^{2}\right)}}+\frac{2 Q^{2}}{l l_{p}} \sqrt{\frac{\left(l^{2}-\eta_0 \beta_0 l_{p}^{2}\right)}{\beta_0}}-\alpha\left(\frac{2}{l l_{p}} \sqrt{\frac{\left(l^{2}-\eta_0 \beta_0 l_{p}^{2}\right)}{\beta_0}}\right)^{3 \omega_{q}}\right].
\end{equation}

\begin{table*}
\begin{tabular}{|c|c|c|c|c||c||c||c||c||c||c|}
\hline
$M_{rem}$ \\
        \hline
\hline
$l=1$& $l_{p}=1$& $Q=0.3$& $\alpha=0.1$\\
\hline
$\omega_{q}=-1/3$ \\

\hline
\hline
$\eta_{0}$ &  $\beta_{0}=0.09$& $\beta_{0}=0.1$ & $\beta_{0}=0.2$ &  $\beta_{0}=0.3$  & $\beta_{0}=0.4$  & $\beta_{0}=0.5$  & $\beta_{0}=0.6$  & $\beta_{0}=0.7$  & $\beta_{0}=0.8$  & $\beta_{0}=0.9$ \\
\hline
\hline
$0.0$ & 0.367500  & 0.355756& 0.301869 & 0.287554 &     0.284605 & 0.286378 &  0.290474 & 0.295819& 0.301869& 0.308322\\
$0.1$& 0.366453  &0.354688  &  0.300868 & 0.286962 &  0.284664  & 0.287289 &0.292410& 0.298942& 0.306328 & 0.314259\\
$0.2$ & 0.365404  & 0.353618& 0.299878 & 0.286421& 0.284852& 0.288453& 0.294783 &0.302750 &0.311800&
0.321627 \\
$0.3$ & 0.364352 & 0.352547  &0.298900 &0.285936  & 0.285187 & 0.289913 & 0.297679 & 0.307407& 0.318566& 0.330884\\
$0.4$&  0.363299  & 0.351473   &0.297935  &0.285514   & 0.285687  & 0.291720& 0.301209 & 0.313130& 0.327023& 0.342712\\
$0.5$& 0.362244 & 0.350398& 0.296985 & 0.285162 & 0.286378&   0.293939&  0.305521 & 0.320220 & 0.337750&
 0.358177\\
$0.6$ & 0.361187 & 0.349322 & 0.296050& 0.284888 & 0.287289 & 0.296649 & 0.310807& 0.329106 & 0.351638
 &0.379063\\
$0.7$& 0.360129& 0.348244 & 0.348244 & 0.284701& 0.288453& 0.299954& 0.317334& 0.340421 &0.370136&
 0.408622\\
$0.8$ & 0.359068 & 0.347164 & 0.294234  &0.284611& 0.289913& 0.303986 &0.325474 &0.355150 & 0.395784&
0.453589\\
$0.9$ &  0.358005 & 0.346083&0.293356 & 0.284631& 0.291720& 0.308922 & 0.335772& 0.374911& 0.433564&
0.531049\\
\hline
$\omega_{q}=-2/3$ \\
\hline
\hline
$0.0$ & 0.373875& 0.362412& 0.310550& 0.297497& 0.295416& 0.297806& 0.302339 & 0.307986 &0.314230&
0.320789\\
$0.1$& 0.372852 & 0.361371& 0.309611& 0.296999& 0.295593& 0.298847& 0.304405 & 0.311223& 0.318771 &
0.326759\\
$0.2$& 0.371826 & 0.360329& 0.308685& 0.296555& 0.295902& 0.300142& 0.306903& 0.315131&
0.324293& 0.334099\\
$0.3$& 0.370800  & 0.359285& 0.307772& 0.296170& 0.296360& 0.301734& 0.309917& 0.319864&
0.331058& 0.343232\\
$0.4$& 0.369771 & 0.358240& 0.306874& 0.295850& 0.296986& 0.303672& 0.313554& 0.325627&
0.339433& 0.354780\\
$0.5$ &0.368741 & 0.357194& 0.305992& 0.295603& 0.297806& 0.306018& 0.317952& 0.332702&
0.349951& 0.369703\\
$0.6$& 0.367709& 0.356146& 0.305128& 0.295437& 0.298847& 0.308849& 0.323294& 0.341484&
0.363416& 0.389576\\
$0.7$& 0.366676 & 0.355097 & 0.304282 & 0.295360 & 0.300142 & 0.312265 & 0.329830 & 0.352553 &
0.381118 & 0.417207\\
$0.8$ &0.365641 & 0.354047& 0.303456& 0.295384& 0.301734& 0.316391& 0.337905& 0.366796&
0.405274& 0.458232\\
$0.9$ &0.364605 &0.352996& 0.302654& 0.295521& 0.303672& 0.321395& 0.348019& 0.385649&
0.440108& 0.526249\\
\hline
\end{tabular}
\caption{the remnant mass as a function for the different EGUP parameters $\beta_{0}$  and $\eta_{0}$.}\label{tabl:em3}
\end{table*}

For $\eta_{0}=0$, GUP modified heat capacity reads
\begin{equation}
C_{G U P}=
\frac{\frac{\pi \beta_0 l_{p}^{2}}{4}\left(1-\frac{Q^{2}}{r_{h}^{2}}+\frac{3 \alpha \omega_{q}}{r_{h}^{3 \omega_{q}+1}}\right) \sqrt{1-\frac{\beta_0 l_{p}^{2}}{4 r_{h}^{2}}}}{ \left(1+\frac{Q^{2}}{r_{h}^{2}}-\frac{9\alpha \omega_q^2}{r_h^{3\omega_q+1}}\right)\left[\sqrt{1-\frac{\beta_0 l_{p}^{2}}{4 r_{h}^{2}}}-1+\frac{\beta_0 l_{p}^{2}}{4 r_{h}^{2}}\right]-\frac{\beta_0 l_{p}^{2}}{4 r_{h}^{2}}\left(1-\frac{Q^{2}}{r_{h}^{2}}+\frac{3 \alpha \omega_{q}}{r_{h}^{3 \omega_{q}+1}}\right)},
\end{equation}
In this scenario, if the horizon satisfies $r_{h}=\frac{l_{p}\sqrt{\beta_0}}{2}$,  then a black hole remnant occurs at the following remnant temperature
\begin{equation}
T_{rem-G U P}=\frac{1}{\pi \sqrt{\beta_0}l_{p}}\left[1-\frac{4 Q^{2}}{\beta_0 l^{2}_{p}}+3 \omega_{q} \alpha \left(\frac{2}{\sqrt{\beta_0}l_{p}}\right)^{3 \omega_{q}+1}\right],
\end{equation}
and remnant mass
\begin{equation}
M_{rem-G U P}=\frac{1}{2}\left[\frac{1}{2} \sqrt{\beta_0}l_{p}+\frac{2 Q^{2}}{\sqrt{\beta_0}l_{p}}-\alpha\left(\frac{2}{\sqrt{\beta_0}l_{p}}\right)^{3 \omega_{q}}\right].
\end{equation}

For $\beta_0=0$, we get EUP modified heat capacity function  in the form of
\begin{equation}
C_{EUP}=\frac{2\pi r_h^2\left(1-\frac{Q^{2}}{r_{h}^{2}}+\frac{3 \alpha \omega_{q}}{r_{h}^{3 \omega_{q}+1}}\right)}{\left(1+\frac{Q^{2}}{r_{h}^{2}}-\frac{9\alpha \omega_q^2}{r_h^{3\omega_q+1}}\right)\bigg(1+\frac{4 \eta_0 r_{h}^{2}}{l^{2}}\bigg)-2\left(1-\frac{Q^{2}}{r_{h}^{2}}+\frac{3 \alpha \omega_{q}}{r_{h}^{3 \omega_{q}+1}}\right)}.
\end{equation}

\begin{figure}
\begin{tabular}{cc}
\begin{minipage}[t]{0.45\linewidth}
\centerline{\includegraphics[width=7.0cm]{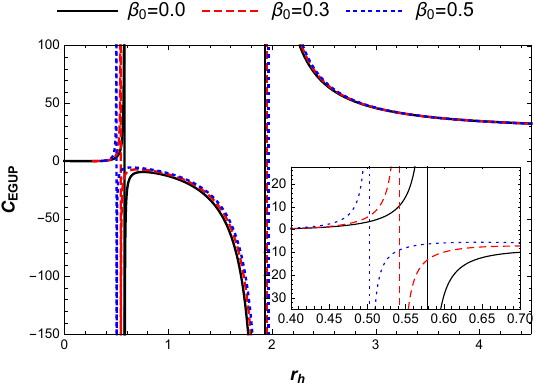}}
\centerline{(a)}
\end{minipage}
\begin{minipage}[t]{0.45\linewidth}
\centerline{\includegraphics[width=7.0cm]{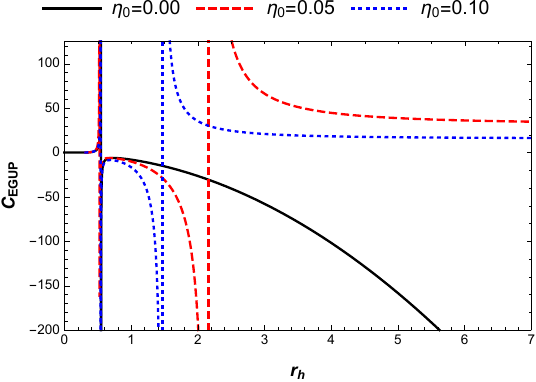}}
\centerline{(b)}
\end{minipage}
\end{tabular}
\parbox[c]{15.0cm}{\footnotesize{\bf Fig.~3.}(a) The heat capacity $C_{EGUP}$ as a function of  horizon radius $r_{h}$ for different GUP parameters $\beta_{0}$ $(l=l_{p}=1, Q=0.3, \omega_{q}=-1/3, \alpha=0.1,\eta_{0}=0.06)$.\\
(b) {The heat capacity $C_{EGUP}$ as a function of horizon radius $r_{h}$ for different EUP parameters $\eta_0$ $(l=l_{p}=1, Q=0.3, \omega_{q}=-1
/3, \alpha=0.1, \beta_0=0.39 )$.}}
\end{figure}
\begin{figure}
\begin{tabular}{cc}
\begin{minipage}[t]{0.45\linewidth}
\centerline{\includegraphics[width=7.0cm]{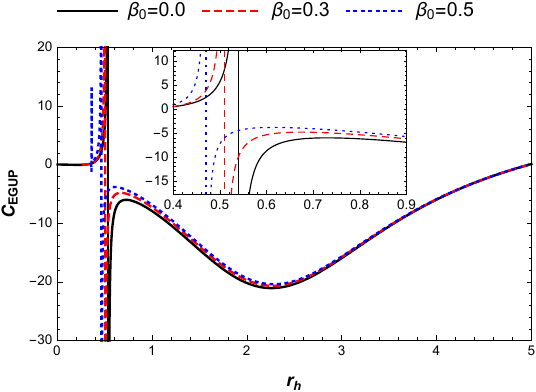}}
\centerline{(a)}
\end{minipage}
\begin{minipage}[t]{0.45\linewidth}
\centerline{\includegraphics[width=7.0cm]{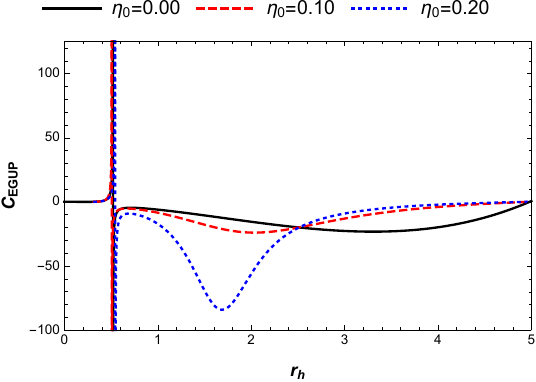}}
\centerline{(b)}
\end{minipage}
\end{tabular}
\parbox[c]{15.0cm}{\footnotesize{\bf Fig.~4.}(a) The heat capacity $C_{EGUP}$ as a function of  horizon radius $r_{h}$ for different GUP parameters $\beta_{0}$ $(l=l_{p}=1, Q=0.3, \omega_{q}=-2/3, \alpha=0.1,\eta_{0}=0.5)$.\\
(b) The heat capacity $C_{EGUP}$ as a function of  horizon radius $r_{h}$ for different EUP parameters $\eta_{0}$ $(l=l_{p}=1, Q=0.3, \omega_{q}=-2/3, \alpha=0.1,\beta_{0}=0.38)$.}
\end{figure}

We observe that in the EUP scenario, a remnant temperature and mass which depends on $\eta_0$, do not exist. In Table \ref{tabl:em3}, we give the effects of the EGUP parameters $\beta_{0}$ and $\eta_{0}$ on the remnant mass under considering the barotropic index as $\omega_{q}=-1/3$ and $\omega_{q}=-2/3$. while for  $\beta_{0}=0$ and $\eta_{0}=0$, the remnant mass first decreases and then increases within the range (0.09, 0.9) of the GUP parameter $\beta_{0}$. {We find that when the dimensionless GUP parameter $\beta_{0}$ is in the range from 0.2 to 0.4,} the effect of the EUP scenario on the remnant mass is almost negligible. However, $\beta_{0}$ in other ranges obviously affects the change of the remnant mass. For the heat capacity, to illustrate this analysis visually, we depict the EGUP modified heat capacity versus horizon for $\omega_q=-1/3$ and $\omega_q=-2/3$ in Fig. 3 and Fig. 4, respectively. In Fig. 3(a), by taking the constant EUP parameter $\eta_{0}= 0.06$, the barotropic index $\omega_{q}=-1/3$, and three GUP parameters, $\beta_{0}=0.0, 0.3, 0.5$, we can observe the trend effect of the different GUP on the heat capacity is very weak. However, for the different EUP parameters, our investigation reveals that the completely different trend of the heat capacity means that the EUP parameter $\eta_{0}$ has a non-negligible effect in Fig. 3, the case of $\omega_{q}=-2/3$ can also be found in Fig. 4.

Next, we use the standard definition to derive the entropy function of the black hole, which is expressed as
\begin{equation}
S=\int\frac{1}{T}dM,
\end{equation}
we combine Eqs. (\ref{t1}) and (\ref{t2}) through simple algebraic derivation, the entropy function is written as

\begin{equation}
S=\frac{\pi l \sqrt{\left(l^{2}-\beta_{0} \eta_{0} l_{p}^{2}\right)} \ln \left[r_{h}\left(2 l^{2}-2 \beta_{0} \eta_{0} l_{p}^{2}+2l \sqrt{\left(l^{2}-\beta_{0} \eta_{0} l_{p}^{2}\right)}\left(\sqrt{\left.1-\frac{\beta_{0} l_{p}^{2}}{4 r_{h}^{2}}\left(1+\frac{4\eta_{0} r_{h}^{2}}{l^{2}}\right)\right)}\right)\right]\right.}{4 \eta_{0}}
+\frac{\pi l^{2} \ln \left(l^{2}+4 \eta_{0} r_{h}^{2}\right)}{8 \eta_{0}}.
\end{equation}
In general, we derive the entropy function of the Reissner-Nordstr\"om black hole under considering the EGUP. If we ignore the EUP ($\eta_{0}=0$), the entropy function reduces to
\begin{equation}
S_{G U P}=\frac{\pi}{2}\left(1+\sqrt{1-\frac{\beta_{0} l_{p}^{2}}{4 r_{h}^{2}}}\right)r^{2}_{h}-\frac{\pi \beta_{0} l_{p}^{2}}{8} \ln (r_{h})-\frac{\pi \beta_{0} l_{p}^{2}}{8} \ln \left(2+2 \sqrt{1-\frac{\beta_{0} l_{p}^{2}}{4 r_{h}^{2}}}\right).
\end{equation}
{If we take the EGUP parameters to be $\eta_0 = \beta_0 = 0$, then the entropy reduces to the usual expression $S=\pi r^{2}_{h}$.}  We can find that QM will not affect the entropy of the black hole. Until now, we have derived the modified  Hawking temperature, heat capacity, and entropy for the Reissner-Nordstr\"om black hole with QM in the EGUP framework.

In the end, we know from Ref \cite{qui1} that the pressure $P_{q}$ and matter-energy density $\rho_{q}$ of the Reissner-Nordstr\"om black hole with QM reads
\begin{equation}
P_{q}=\omega_{\mathrm{q}} \rho_{q}=-\frac{3}{2} \frac{\alpha \omega_{q}^{2}}{r^{3\left(\omega_{\mathrm{q}}+1\right)}}.
\end{equation}
Based on this, we consider the relations between the event horizon in Eq. (\ref{t1}) and the matter-energy density, the mass in terms of the pressure and horizon radius is given by
\begin{equation}
M=\frac{1}{2}\left(r_{h}+\frac{Q^{2}}{r_{h}}+\frac{P_{q} r_{h}^{3}}{3 \omega^{2}_{q}}\right).
\end{equation}

In this case, the volume $V=\frac{ r_{h}^{3}}{3 \omega^{2}_{q}}$ in terms of the radius \cite{k11}, the Hawking temperature in Eq. (\ref{Te1}) can be re-expressed as
\begin{equation}
T=\frac{\left(3 \omega_{q}^{2} V\right)^{1 / 3}}{\pi \beta_{0} l^{2}_{p}}\left(1-\frac{Q^{2}}{\left(3 \omega_{q}^{2} V\right)^{2 / 3}}-\frac{2 P_{q}\left(3 \omega_{q}^{2} V\right)^{2 / 3}}{\omega_{q}}\right)\left(1-\sqrt{1-\frac{\beta_{0} l^{2}_{p}}{4\left(3 \omega_{q}^{2} V\right)^{2 / 3}}\left(1+\frac{4 \eta_{0}\left(3 \omega_{q}^{2} V\right)^{2 / 3}}{l^{2}}\right)}\right).
\end{equation}
For T=1 isotherm, the EGUP-corrected equation of state of the black hole is given by
\begin{equation}
P_{q}=\frac{\omega_{q}}{2\left(3 \omega_{q}^{2} V\right)^{2 / 3}}\left[1-\frac{Q^{2}}{\left(3 \omega_{q}^{2} V\right)^{2 / 3}}-\frac{\pi \beta_{0} l^{2}_{p}}{\left(3 \omega_{q}^{2} V\right)^{1 / 3}} \frac{1}{\left.1-\sqrt{1-\frac{\beta_{0} l^{2}_{p}}{4\left(3 \omega_{q}^{2} V\right)^{2 / 3}}\left(1+\frac{4 \eta_{0}\left(3 \omega_{q}^{2} V\right)^{2 / 3}}{l^{2}}\right)}\right)}\right].
\end{equation}
Now, we have derived the EGUP-corrected Pressure-Volume (P-V) isotherm for the black hole with QM. For $\eta_{0}=0$, the GUP-corrected pressure of black hole reads
\begin{equation}
P_{q-G U P}=\frac{\omega_{q}}{2\left(3 \omega_{q}^{2} V\right)^{2 / 3}}\left[1-\frac{Q^{2}}{\left(3 \omega_{q}^{2} V\right)^{2 / 3}}-\frac{\pi \beta_{0} l^{2}_{p}}{\left(3 \omega_{q}^{2} V\right)^{1 / 3}} \frac{1}{\left(1-\sqrt{1-\frac{\beta_{0} l^{2}_{p}}{4\left(3 \omega_{q}^{2} V\right)^{2 / 3}}}\right)}\right] .
\end{equation}
For $\beta_{0}=0$, the EUP-corrected pressure of black hole reads
\begin{equation}
P_{q-E U P}=\frac{\omega_{q}}{2\left(3 \omega_{q}^{2} V\right)^{2 / 3}}\left[1-\frac{Q^{2}}{\left(3 \omega_{q}^{2} V\right)^{2 / 3}}-\frac{4 \pi\left(3 \omega_{q}^{2} V\right)^{1 / 3}}{\left(1+\frac{4 \eta_{0}\left(3 \omega_{q}^{2} V\right)^{2 / 3}}{l^{2}}\right)}\right].
\end{equation}
while for $\beta_{0}=0$ and $\eta_{0}=0$, Eq. (42) reduces to HUP-modified pressure

\begin{equation}
P_{q-H U P}=\frac{\omega_{q}}{2\left(3 \omega_{q}^{2} V\right)^{2 / 3}}\left[1-\frac{Q^{2}}{\left(3 \omega_{q}^{2} V\right)^{2 / 3}}-4 \pi\left(3 \omega_{q}^{2} V\right)^{1 / 3}\right].
\end{equation}
\begin{figure}
\begin{tabular}{cc}
\begin{minipage}[t]{0.45\linewidth}
\centerline{\includegraphics[width=7.0cm]{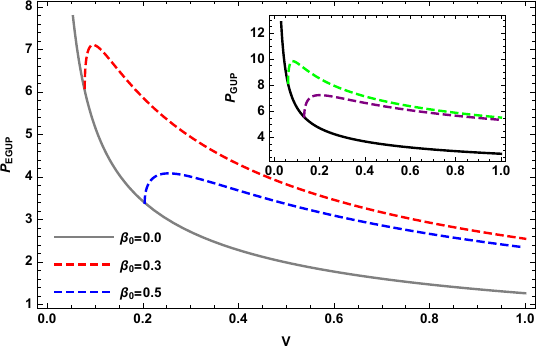}}
\centerline{(a)}
\end{minipage}
\begin{minipage}[t]{0.45\linewidth}
\centerline{\includegraphics[width=7.0cm]{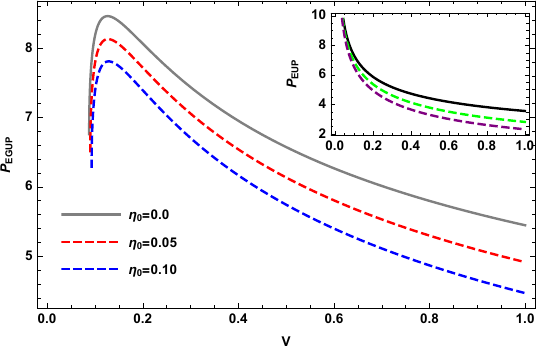}}
\centerline{(b)}
\end{minipage}
\end{tabular}
\parbox[c]{15.0cm}{\footnotesize{\bf Fig.~5.}(a) The EGUP-corrected pressure as a function of the volume V for different GUP parameter values  $\beta_{0}$. $(l=l_{p}=1, Q=0.3, \omega_{q}=-1/3, \alpha=0.1, \eta_0=0.5)$  \\
(b) The EGUP-corrected pressure as a function of the volume V for different EUP parameter values $\eta_0$. $(l=l_{p}=1, Q=0.3, \omega_{q}=-1/3, \alpha=0.1, \beta_0=0.38)$. }
\end{figure}

\begin{figure}
\begin{tabular}{cc}
\begin{minipage}[t]{0.45\linewidth}
\centerline{\includegraphics[width=7.0cm]{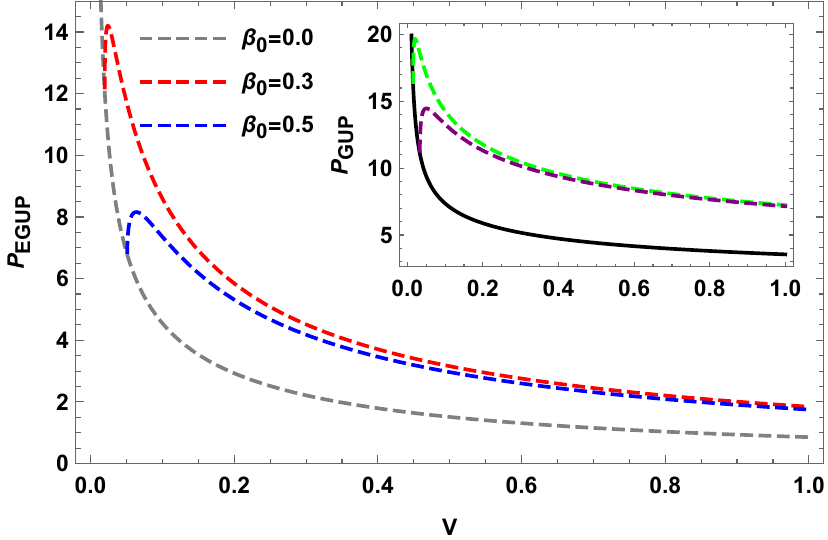}}
\centerline{(a)}
\end{minipage}
\begin{minipage}[t]{0.45\linewidth}
\centerline{\includegraphics[width=7.0cm]{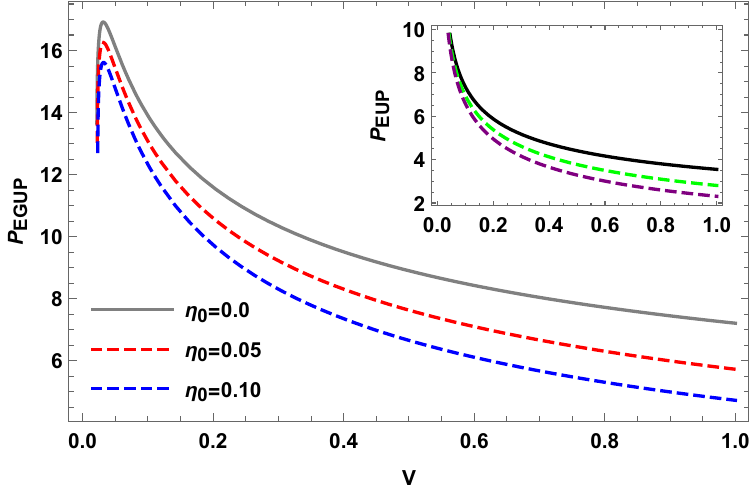}}
\centerline{(b)}
\end{minipage}
\end{tabular}
\parbox[c]{15.0cm}{\footnotesize{\bf Fig.~6.}(a)The EGUP-corrected pressure as a function of the volume V for different GUP parameter values  $\beta_{0}$. $(l=l_{p}=1, Q=0.3, \omega_{q}=-2/3, \alpha=0.1, \eta_0=0.5)$  \\
(b) The EGUP-corrected pressure as a function of the volume V for different EUP parameter values $\eta_0$. $(l=l_{p}=1, Q=0.3, \omega_{q}=-2/3, \alpha=0.1, \beta_0=0.38)$.}
\end{figure}

To analyze the effect of EGUP parameters $\beta_0, \eta_0$ on the P-V isotherm, we depict the modified P-V isotherm for the barotropic index $\omega_{q}=-1/3$ and $\omega_{q}=-2/3$ in Fig. 5 and Fig.6, respectively. To this end, we consider the parameter values $(l=l_{p}=1, Q=0.3,\alpha=0.1)$, we can find that the EGUP modified P-V isotherm  have similar trends (blue and red dashed line) when both EUP $\eta_0$ and GUP $\beta_0$ are not zero, the EUP $\eta_0$ is weak in contrast to HUP (black line).  However, GUP has a crucial effect on the P-V isotherm.
\section{Conclusions} \label{sec:conclusion}

In this work, we use the extended generalized uncertainty principle to study the thermodynamics of the Reissner-Nordstr\"om black hole with QM. First of all, the different modified uncertainty principles are briefly reviewed, based on the concept of the minimum position and momentum, the extended generalized uncertainty principle is selected. Afterward, the quantum corrected-Hawking temperature of the Reissner-Nordstr\"om black hole with QM is obtained. To better understand the concept of the accelerated expansion of the universe and quintessence regime of dark energy, we take the barotropic index $\omega_{q}$ as $-1/3$ and $-2/3$. On the one hand, the different values of the barotropic index $\omega_{q}$ will result in different areas of  the effective horizon radius, the EGUP parameters $\beta_{0}$ and $\eta_{0}$ lead to a lower bound on the horizon. On the other hand, to analyze and demonstrate the influence of the GUP and EUP parameters, the characteristic of the horizon range and Hawking temperature in Tables \ref{tab:tab0} and \ref{tab:tab1} are summarized in conjunction with Fig. 1 and Fig. 2. In the next part, we study the modified remnant mass, heat capacity, and entropy functions of the considered black hole. It can be observed that the values of the GUP parameter $\beta_{0}$ in the range from 0.2 to 0.4, the effect of the EUP scenario on the remnant mass is almost negligible in Table \ref{tabl:em3}. {Note that the quintessence matter does not affect the black hole's entropy.} In the end, a detailed analysis of  the EGUP modified P-V isotherm with graphical analysis is presented.

\section*{Acknowledgments}
The authors are grateful to anonymous reviewers for their very crucial and detailed comments. The authors would like to thank Prof. Ali \"Ovg\"un from the Physics Department, Eastern Mediterranean University for his valuable comments and suggestions during the completion of this manuscript. The author Z.W. Long is supported by the National Natural Science Foundation of China (Grant nos. 11465006 and 11565009) and the Major Research Project of innovative Group of Guizhou province (2018-013). B. C. L\"{u}tf\"{u}o\u{g}lu is supported by the Internal  Project,  [2022/2218],  of  Excellent  Research  of  the  Faculty  of  Science  of  University.





\end{document}